\documentclass[prl,aps,twocolumn,showpacs,amsmath,amssymb]{revtex4}
\usepackage{graphicx}
\usepackage{dcolumn}

\begin{document}

\title{Four-Probe Measurements of Carbon Nanotubes with Narrow Metal Contacts}

\author{A. Makarovski$^{1}$, A. Zhukov$^{1}$, J. Liu$^{2}$, and G. Finkelstein$^{1}$}
\affiliation{Departments of $^1$Physics and $^2$Chemistry, Duke University, Durham, NC 27708}

\begin{abstract}

We find that electrons in single-wall carbon nanotubes may propagate substantial distances (tens of nanometers) under the metal contacts. We perform four-probe transport measurements of the nanotube conductance and observe significant deviations from the standard Kirchhoff's circuit rules. Most noticeably, injecting current between two neighboring contacts on one end of the nanotube, induces a non-zero voltage difference between two contacts on the other end. 

\end{abstract}

\pacs{PACS numbers: 73.23.-b, 73.21.Hb, 73.63.Rt, 73.63.Fg }



\maketitle


Multiprobe transport measurements have been at the heart of the developments in mesoscopic physics \cite{Imry}. However, most of the conductance measurements on single-wall carbon nanotubes are nowadays performed in a two-probe geometry, where the same leads are used to supply voltage and measure current (or vice versa). Such measurements are necessarily affected by the properties of the nanotube-metal contacts, which are essentially connected in series with the nanotube itself.  Multi-probe measurements do not provide a significant advantage, since the metal contacts separate the nanotube into segments \cite{Bezryadin1998,Mann2003}, and the electrons fully equilibrate in the metal upon leaving each segment. Therefore, the results of the 2- and 4-probe measurements usually coincide. In a notable exception, in Ref. \cite{Gao2005} potential (voltage) probes to a nanotube have been made of multi-wall nanotubes, which allow for a true 4-probe measurement. Also, a scanning microscope tip may serve as a weakly coupled potential probe that does not significantly disturb the electron flow in the nanotube beneath \cite{Bachtold2000,Yaish2004}.

In this work, we investigate the nanotube resistances measured in a four-probe set-up with metal contacts (image in Figure 1). The two central electrodes are made sufficiently narrow ($\sim 40$ nm), which allows some fraction of electrons to stay in the nanotube while traversing the electrode. The electronic transport in this situation can be rather non-intuitive ({\it i.e.} it may contradict the standard circuit rules). Most interestingly, when current flows through two neighboring contacts on one end of the nanotube, we find a non-zero voltage difference between the two contacts on the other end. This {\it non-local} four-probe measurement allows us to study mode equilibration in the nanotube.

The single-wall carbon nanotubes were grown by a CVD method using CO as a feedstock gas (the details are described in our earlier publication \cite{Zheng2002}). PdAu contacts are patterned by e-beam lithography and deposited by thermal evaporation on top of the nanotube. We measure the metallic nanotube transport properties by supplying a fixed AC current (supplied through a 10 $M \Omega$ resistor) and measuring the resulting AC voltage at temperatures down to 1.3 K. Low currents (1-10 nA) are used at lower temperatures, while at higher temperatures (Figures 4, 5) the current is boosted to 100 nA.  The frequency of the excitation signal is kept below 100 Hz, which was verified to be low enough to avoid spurious pick-up signals. We choose to present here results measured on the nanotube imaged in Figure 1. The lengths of the three nanotube segments are 400, 200, and 400 nm, and the two middle electrodes are 40 nm wide. We denote the four contacts to the nanotube A, B, C, and D, as labeled in the schematic in Figure 1. 


\begin{figure}[h]
\includegraphics[width=1.0\columnwidth]{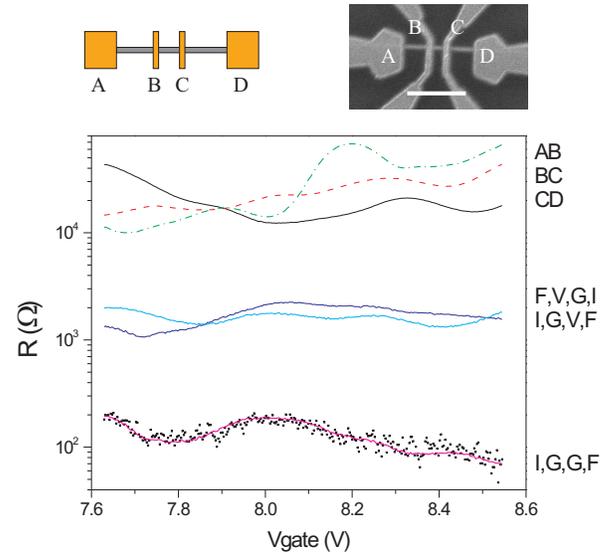}
\caption{\label{fig:segments} Top: the schematic and the scanning electron micrograph of the sample. White bar: 1 $\mu m$. Main panel, top curves: Differential resistances of the 3 nanotube segments measured by a 2-probe method between pairs of contacts AB, BC and CD. Middle curves: the 3-terminal contact resistances $R_{i,g,v,f}$ and $R_{f,v,g,i}$ (see clarification in the text). Bottom: $R_{i,g,g,v}$ data (dots) and the fit described in the text. T=1.3 K. }
\end{figure}

The topmost curves in Figure 1 correspond to two-probe resistances measured between pairs of contacts ($R_{AB}$, $R_{BC}$ and $R_{CD}$), while the two other contacts are floating. The resistances in the range of a few tens of $k\Omega$ should be compared to the minimal resistance of $h/4 e^2\approx 6.5 k\Omega$, possible for ballistic single-wall nanotubes with ideal contacts (no reflections). The smooth oscillatory variations of resistance with gate voltage are due to interference of standing waves in the nanotube. In the following, we measure the nanotube resistance in various contact configurations as a function of the gate voltage $V_{gate}$. Owing to the unique resistance patterns, we can determine contributions of different parts of the structure to the measured signal.


The middle curves in Figure 1 represent the 3-terminal resistances \cite{Buttiker1988}. $R_{i,g,v,f}$ is measured when a fixed current is supplied through A, contact B is grounded, voltage is sensed at C, and D is floating. $R_{f,v,g,i}$ denotes an inverse configuration, where the current is supplied through D, contact C is grounded, and voltage is sensed at B, while A is floating. (We have to depart from the usual notation of Ref. \cite{Buttiker1986,Buttiker1988} which is not applicable to some of our measurements.)  Conventionally, one may expect $R_{i,g,v,f}$ to be a $V_{gate}$-independent constant, corresponding to the metal lead resistance ($R_0 \lesssim 500\Omega$). This would be the case if the electrons arriving from A were completely equilibrated at the metal of contact B. Then potential at C would be equal to potential at B and given simply by $R_0 I$. The excess voltage that we measure at C indicates that some of the electrons injected from A go past contact B {\it staying} in the nanotube channel. Arriving at C, these electrons raise its potential. This excess voltage appears in $R_{i,g,v,f}$ as an apparent excess resistance on top of $R_0$. 

From the value of $R_{i,g,v,f}$, one may estimate the fraction of electrons that go past contact B without equillibration: $\alpha_B \approx (R_{i,g,v,f}-R_0)G_0 \sim $0.1 --- 0.3, where $G_0=4e^2/h$. One can understand this expression as follows: If current $I$ is injected into contact A, its fraction $\alpha_B I$ goes past contact B. The fraction of this current that reaches C is determined by the transparency of the middle section of the nanotube, which according to the Landauer formula is equal to $1/(G_0 R_{BC})$. The resultant current $\alpha_B I/(G_0 R_{BC})$ has to be compensated by an equal but oppositely directed current emerging due the voltage difference between contacts B and C: $\alpha_B I/(G_0 R_{BC})\approx (V_C-V_B)/R_{BC} = (IR_{i,g,v,f}-IR_0)/R_{BC}$. In the following, we ignore $R_0$ compared to $R_{i,g,v,f}$ and $R_{f,v,g,i}$.

The 3-terminal measurements of nanotubes with narrow metal contacts are theoretically discussed in Ref. \cite{Ke2006}. The values of $\alpha \lesssim 0.5$ are found for the metal contacts $\sim 2$ nm wide. One may expect $\alpha$ to decay exponentially with the contact width. However, we observe nonvanishing values of $\alpha$ in our much wider contacts ($\sim 40$ nm). Apparently, in our case the metal is not as effective in wetting the nanotube as it is assumed in the model simulation.

We have checked that at zero magnetic field the 3-terminal resistances do not change upon permutation of the current and voltage leads: $R_{i,g,v,f} = R_{v,g,i,f}$ and $R_{f,v,g,i} = R_{f,i,g,v}$, as it should be according to the Onsager relations \cite{Buttiker1986,Buttiker1988}. When magnetic field parallel to the nanotube axis is applied, the three-terminal resistances of the same contact measured in two configurations ({\it e.g} $R_{i,g,v,f}$ and $R_{v,g,i,f}$) become distinctly different. 


The scattered dots at the bottom of Figure 1 correspond to the measurement configuration denoted $R_{i,g,g,v}$, where a fixed current is supplied through A, both B and C are grounded, and voltage is sensed at D. The reverse configuration is denoted $R_{v,g,g,i}$. Again, these two quantities are found to be equal, as expected from the Onsager relations (one may view the two central grounded electrodes as one terminal). 


We may estimate the current which crosses the two grounded electrodes and the central nanotube segment, as  $\approx I \alpha_B \alpha_C /(R_{BC}G_0)$. We substitute
$\alpha_B \approx R_{i,g,v,f}/G_0$ and $\alpha_C \approx R_{f,v,g,i}/G_0$ to get the current of
$\approx G_0 R_{i,g,v,f} R_{f,v,g,i}/R_{BC}$. The voltage on D required to compensate for this current is $\approx V R_{i,g,v,f} R_{f,v,g,i}/R_{BC}$. The lowest curve in Figure 1 (superimposed over the $R_{i,g,g,v}$ data) shows the resulting formula $R_{i,g,g,v} = R_{i,g,v,f} R_{f,v,g,i}/R_{BC}$, where all the parameters are taken from the previous measurements. 

\begin{figure}
\includegraphics[width=0.80\columnwidth]{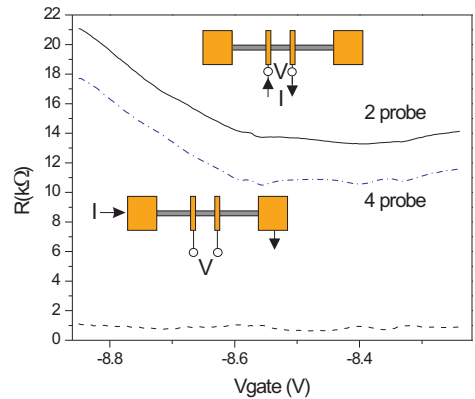}
\caption{\label{fig:R additions} Differential resistance of the middle segment measured by a 2-probe method (solid line) and by a 4-probe method (dash-dotted line). Dashed line: $R_{BC}^{(2)}-R_{BC}^{(4)}-R_{i,g,v,f}-R_{f,v,g,i}$. T=1.3 K.}
\end{figure}


The good quality of the fit indicates that our simple considerations capture the essential features of the system. Nonetheless, an essentially identical formula may be obtained if we formally apply Kirchhoff's circuit rules, by artificially taking $R_{i,g,v,f}$ and $R_{f,v,g,i}$ for the contact resistances and assuming $R_{BC}$ is much larger then either of them. In this case, the nanotube would work as a voltage divider, and we would arrive at the same expression for the measured voltage. To clearly demonstrate that the conventional intuition based on the resistor circuit rules does not work in our structure, we present in Figure 2 the two-probe resistance of the middle segment (denoted $R_{BC}^{(2)}$) and the resistance of the same segment measured in a conventional four-probe scheme, where current is supplied to A, D is grounded, and the voltage difference is measured between B and C (denoted $R_{BC}^{(4)}$). Naively, the two measurements should differ by the resistances of the contacts B and C (including the resistances of the metal-nanotube interfaces, as in the 3-terminal arrangement in Figure 1).  The lower curve in Figure 1b shows $R_{BC}^{(2)}-R_{BC}^{(4)}-R_{i,g,v,f}-R_{f,v,g,i}$. As we see, the four-probe resistance of the central segment plus the resistances of the contacts do not add up to the total two-probe resistance. This indicates that the Kirchhoff's circuit rules are inadequate for describing the transport, and should be replaced by the Landauer- Buttiker formalism \cite{Imry,Buttiker1986,Buttiker1988}. Indeed, if the electrons flowing in the nanotube only partially equilibrate with the narrow central electrodes, the four-probe resistance stops to reflect the actual resistivity \cite{Timp1988}, and may even become negative \cite{dePicciotto2001,Gao2005}. 

\begin{figure}
\includegraphics[width=0.80\columnwidth]{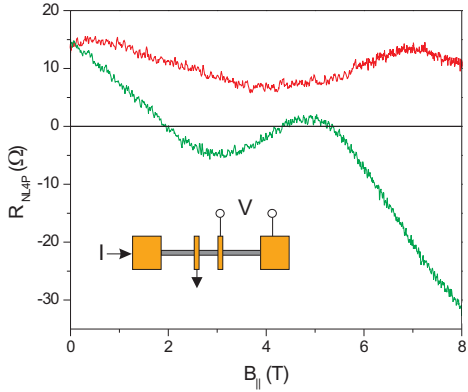}
\caption{\label{fig:NL4P B} Nonlocal four-probe resistance measurements $R_{i,g,v-,v+}$ (see schematic) and $R_{v+,v-,g,i}$ as a function of magnetic field parallel to the nanotube. T=1.3 K.}
\end{figure}

In the rest of the paper, we consider a non-local four-probe (NL4P) measurement, where a fixed current is supplied through A, contact B is grounded, and a voltage difference is measured between C and D (denoted $R_{i,g,v-,v+}$). Similar nonlocal ``bend resistance'' measurements were performed in ballistic GaAs channels \cite{Timp1988}. Typically, this signal is very small, an order of magnitude smaller than $R_{i,g,g,v}$ discussed previously. NL4P should vanish within the Landauer theory if the nanostructure has only one mode (see Figure 12 in \cite{Baranger1990}): since no current is flowing into C and D, the voltage drop between them should be identically zero. However, we have to recall that nanotubes have 2 transversal modes. If these modes have different transmission coefficients and are differently coupled to the contacts, then nonvanishing NL4P signal $R_{i,g,v-,v+}$ may appear. 

Figure 3 shows the two NL4P signals ($R_{i,g,v-,v+}$ and $R_{v+,v-,g,i}$) measured as a function of magnetic field parallel to the nanotube at a fixed gate voltage. The two signals coincide at zero field, as it should be according to Onsager relations. We see that the two measurements split when the time-reversal symmetry is broken by the magnetic field. A substantial difference between the two is accumulated already in a field of $\sim 1$T. Indeed, the parallel field couples strongly to the orbital motion of the electrons around the circumference of the nanotube, splitting the two propagating modes. 

NL4P was studied previously in multi-walled nanotubes in Ref. \cite{Bourlon2004}, where it was used to investigate the conductance between the concentric shells. In those experiments, some of the current flows in the inner shell past the drain terminal toward the potential probes, and then flows back toward the drain in the outer shell. As a result, the voltage probe closer to the drain always acquires a lower potential than the probe farther away. In our measurements, we observe NL4P voltage of both signs. To qualitatively illustrate the possibility of such behavior, one can imagine current flowing from B only in mode (1), and returning from D only in mode (2). The relative potential of contact C will be different if it couples preferentially to mode (1) or mode (2). 

\begin{figure}
\includegraphics[width=1\columnwidth]{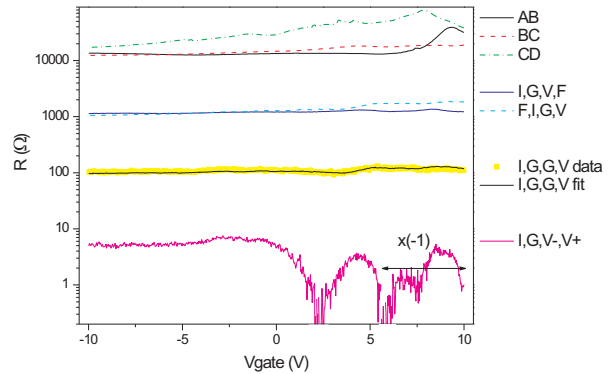}
\caption{\label{fig:all 70k} Various signals discussed in the paper measured at the temperature of 70 K. Top 3 curves ($R \gtrsim 10 k\Omega$): segment resistances. Next lower curves $R \gtrsim 1 k\Omega$: 3-terminal resistances. Scattered squares with overlayed curve ($R \sim 100 \Omega$): $R_{i,g,g,v}$ measurement and the corresponding fit. Lowest curve: nonlocal four-probe resistance. Note that the NL4P signal changes sign at $V_{gate} \approx 5.6$ V.}
\end{figure}

To be more specific, we use the 4-probe formalism of Refs. \cite{Buttiker1986,Buttiker1988} to express $R_{i,g,v-,v+}$ as
$\frac{h}{e^2}(T_{AD} T_{BC}-T_{AC}T_{BD})/{\cal D}$, where $T_{AC}$ indicates the electron transmission between A and C while B and D are grounded, and similarly for $T_{BD}$ and $T_{AD}$. Here ${\cal D}$ is a denominator defined in \cite{Buttiker1986,Buttiker1988}. In our case ${\cal D}\approx T_{AB} T_{BC} T_{CD}$. Throughout the text, we have tacitly assumed that $T_{AC} = \alpha_B T_{AB}T_{BC}$, $T_{BD} = \alpha_C T_{BC} T_{CD}$ and $T_{AD} = \alpha_B \alpha_C T_{AB} T_{BC} T_{CD}$, in which case  $R_{i,g,v-,v+}$ should be identically equal to zero, as $T_{AC} T_{BD} = T_{AD} T_{BC}$. However, let us now take into account the presence of two modes, so that $T_{AB} =T_{AB}^{(1)}+T_{AB}^{(2)}$, $\alpha_A = \alpha_A^{(1)}+\alpha_A^{(2)}$ and similarly for $T_{BC}$, $T_{CD}$ and $\alpha_B$. Here superscripts (1) and (2) indicate the mode index. Let us assume for simplicity that the modes do not mix under the metal contacts, so that  $T_{AC}=\alpha_B^{(1)}T_{AB}^{(1)}T_{BC}^{(1)}+\alpha_B^{(2)}T_{AB}^{(2)}T_{BC}^{(2)}$ and similarly for 
$T_{BD}$ and $T_{AD}$. A straightforward algebra leads to the following expression:$R_{i,g,v-,v+}= \frac{h}{e^2}T_{BC}^{(1)}T_{BC}^{(2)}(T_{AB}^{(1)}\alpha_B^{(1)}-T_{AB}^{(2)}\alpha_B^{(2)})(T_{CD}^{(1)}\alpha_C^{(1)}-T_{CD}^{(2)}\alpha_C^{(2)})/{\cal D}$. It is now enough to have  $T_{AB}^{(1)}\alpha_B^{(1)} \neq T_{AB}^{(2)}\alpha_B^{(2)}$ and $T_{CD}^{(1)}\alpha_C^{(1)} \neq T_{CD}^{(2)}\alpha_C^{(2)}$ to ensure that $R_{i,g,v-,v+} \neq 0$. We expect these inequalities to be naturally satisfied in real nanotubes, where the mode splitting should be substantial.


\begin{figure}
\includegraphics[width=0.80\columnwidth]{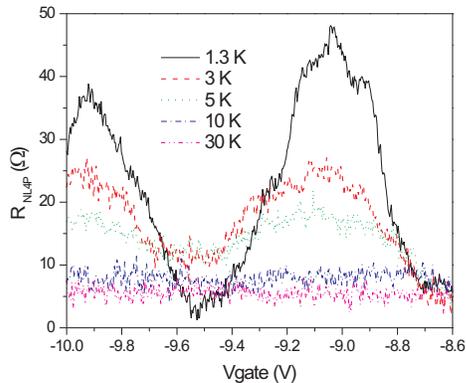}
\caption{\label{fig:NL4P T} Nonlocal four-probe resistance as a function of gate voltage at several temperatures. The modulations of the NL4P signal decay rapidly with temperature. However, nonzero signal is still left at the highest temperature. }
\end{figure}

Finally, in Figure 4, we present all the signals discussed in the paper, measured at $T=70$ K. Most of the observations made in this paper at $T= 1.3$ K still apply here. These include the noticeable three-terminal contact resistance and the non-local four-probe resistance. As mentioned earlier, NL4P signal of either sign is observed. The magnitude of this signal at 70 K (typically Ohms) is noticeably smaller than that observed at 1.3 K (typically tens of Ohms). As we discussed above, the NL4P is sensitive to the presence of two modes, which have different transmission coefficients. One can intuitively argue that the elevated temperature should increase the mode equilibration in the nanotube, reducing the effect of the differences between their transmission coefficients. (Full mode equilibration is essentially identical to having only one mode.)

In Figure 5 we demonstrate the evolution of NL4P with temperature. Clearly, the oscillations in the NL4P signal rapidly decay with temperature, although the leftover signal of $\sim 5 \Omega$ survives. The disappearance of the oscillations in NL4P correlates with the flattening of the resistance curves for individual segments (Figure 4, top curves, $V_{gate} < 0$). It therefore simply reflects the smearing of the single particle interference due to energy spread of the participating electrons. Interestingly, the  NL4P signal does not vanish even at 70 K. This indicates that the population equilibration between the two modes is still not very effective at this relatively high temperature.


In conclusion, we report on multi-probe measurements of single-wall carbon nanotubes with narrow metal electrodes. We estimate the fraction of electrons which pass across the $40$ nm electrodes without equilibration as $\alpha \gtrsim 0.1$. We show how Kirchhoff's circuit rules break down due to electrons in the nanotube flowing across narrow electrodes without full equilibration. We study the non-local four-probe measurement, which directly reflects the presence of more than one transport mode in the nanotube. Our results indicate that the mode equilibration is not complete even at the temperature of 70 K. 


Acknowledgements: We thank H. Baranger and S. Teitsworth for valuable discussions. The work is supported by NSF DMR-0239748.

\end{document}